\documentclass[aps,10pt,twocolumn,superscriptaddress]{revtex4}

\usepackage{amssymb,amsmath}
\usepackage{subfig}
\usepackage{graphicx}
\usepackage{color}
\usepackage{hyperref}
\usepackage{listings}
\usepackage{ragged2e}
\usepackage{tabu}

\hypersetup{
    bookmarks=true,         
    unicode=false,          
    pdftoolbar=true,        
    pdfmenubar=true,        
    pdffitwindow=false,     
    pdfstartview={FitH},    
    pdfauthor={Armaos, Argiriou, Kioutsioukis},     
    colorlinks=true,       
    linkcolor=blue,          
    citecolor=blue,        
}
\graphicspath{{figures/}}

\begin{document}

\definecolor{dkgreen}{rgb}{0,0.6,0}
\definecolor{gray}{rgb}{0.5,0.5,0.5}
\definecolor{mauve}{rgb}{0.58,0,0.82}

\captionsetup{justification=justified,singlelinecheck=false,labelfont=large}

\lstset{frame=tb,
  	language=Matlab,
  	aboveskip=3mm,
  	belowskip=3mm,
  	showstringspaces=false,
  	columns=flexible,
  	basicstyle={\small\ttfamily},
  	numbers=none,
  	numberstyle=\tiny\color{gray},
 	keywordstyle=\color{blue},
	commentstyle=\color{dkgreen},
  	stringstyle=\color{mauve},
  	breaklines=true,
  	breakatwhitespace=true
  	tabsize=3
}

\title{Quantum Computing and Atmospheric Dynamics: Exploring the Lorenz System}

\author{V. Armaos}
\affiliation{Laboratory of Atmospheric Physics, Department of Physics, University of Patras, Greece}

\author{Athanassios A. Argiriou}
\affiliation{Laboratory of Atmospheric Physics, Department of Physics, University of Patras, Greece}

\author{Ioannis Kioutsioukis}
\affiliation{Laboratory of Atmospheric Physics, Department of Physics, University of Patras, Greece}

\date{\today}

\begin{abstract}

    This paper explores the potential contribution of quantum computing, specifically the Variational Quantum Eigensolver (VQE), into atmospheric physics research and application problems using as an example the Lorenz system, a paradigm of chaotic behavior in atmospheric dynamics. Traditionally, the complexity and non-linearity of atmospheric systems have presented significant computational challenges. However, the advent of quantum computing, and in particular the VQE algorithm, offers a novel approach to these problems. The VQE, known for its efficiency in quantum chemistry for determining ground state energies, is adapted in our study to analyze the non-Hermitian Jacobian matrix of the Lorenz system. We employ a method of Hermitianization and dimensionality augmentation to make the Jacobian amenable to quantum computational techniques. This study demonstrates the application of VQE in calculating the eigenvalues of the Lorenz system's Jacobian, thus providing insights into the system's stability at various equilibrium points. Our results reveal the VQE's potential in addressing complex systems in atmospheric physics. Furthermore, we discuss the broader implications of VQE in handling non-Hermitian matrices, extending its utility to operations like diagonalization and Singular Value Decomposition (SVD), thereby highlighting its versatility across various scientific fields. This research extends beyond the realm of chaotic systems in atmospheric physics, underscoring the significant potential of quantum computing to tackle complex, real-world challenges.
  
\end{abstract}

\maketitle

\section{\label{Introduction}Introduction}
    Atmospheric physics has long presented computational challenges due to the nonlinear and chaotic nature of atmospheric systems. One such system, the Lorenz system, stands as a classic example of chaos theory, elucidating the unpredictable behavior of weather systems and the inherent limitations of long-term weather forecasting.
    
    The advent of quantum computing offers a new paradigm for tackling such complex systems. Quantum computing, fundamentally different from classical computing in its approach to data processing, harnesses the principles of quantum mechanics to perform calculations. This emerging technology has shown significant promise in various fields, including cryptography, optimization problems, and, most pertinently for our purposes, in simulating quantum mechanical systems \cite{McArdle2020, Armaos2020}.
    
    The Lorenz system, a set of deterministic differential equations first formulated by Edward Lorenz in 1963, provides a simplified mathematical model for atmospheric convection. It is characterized by its chaotic solutions for certain parameter values and initial conditions, epitomizing the "butterfly effect" in chaos theory. The system's sensitivity to initial conditions highlights the challenges of producing accurate long-term weather forecasts, making it a prime candidate for exploring the capabilities of quantum computing in atmospheric physics. \cite{Tennie2023}
    
    This paper aims to delve into the integration of quantum computing, specifically the Variational Quantum Eigensolver (VQE) algorithm, into the study of the Lorenz system within atmospheric dynamics. The VQE, a hybrid quantum-classical algorithm, has demonstrated efficiency in determining the ground state energies of molecular systems, making it well-suited for use in noisy intermediate-scale quantum (NISQ) devices \cite{Peruzzo2014, McClean2016}. By applying the VQE algorithm to the Lorenz system, we aim to explore the potential advantages of quantum computing over classical methods in the analysis of chaotic systems.
    
    The scope of this paper encompasses a thorough examination of the Lorenz system using quantum computational techniques, focusing on the linearization of its differential equations and subsequent analysis using the VQE. Our objective is not only to contribute to the field of atmospheric physics by providing a novel approach to studying chaotic systems but also to expand the applicability of quantum computing in solving complex, real-world problems.

\section{\label{TheoreticalFramework}Theoretical Framework}
    \subsection{\label{LorenzSystem}The Lorenz System}
        The Lorenz system, a set of ordinary differential equations, is pivotal in atmospheric physics, representing a simplified model of atmospheric convection. First formulated by Edward Lorenz in 1963, this system is renowned for exhibiting chaotic behavior, thus laying the groundwork for the study of deterministic chaos in natural systems \cite{Lorentz1963}. Its equations are given by:

        \begin{align}
            \frac{dx}{dt} &= \sigma(y - x), \nonumber \\
            \frac{dy}{dt} &= x(\rho - z) - y, \\
            \frac{dz}{dt} &= xy - \beta z, \nonumber
        \end{align}

        where $\sigma$, $\rho$, and $\beta$ are positive system parameters representing the physical characteristics of the atmosphere. Specifically, $\sigma$ is the Prandtl number, $\rho$ is the Rayleigh number, and $\beta$ is a proportionality factor. The Lorenz system is nonlinear, with its dynamics fundamentally influenced by these parameters.

        The equilibrium points of the Lorenz system are determined by setting the derivatives to zero, leading to the equations:

        \begin{align}
            \sigma(y - x) &= 0, \nonumber \\
            x(\rho - z) - y &= 0, \\
            xy - \beta z &= 0. \nonumber
        \end{align}

        Solving these equations, the equilibrium points are identified as

        \begin{align}
            \left. \begin{array}{l}
                \phantom{(x_0, y_0, z_0) =} \\
                (x_0, y_0, z_0) = \\
                \phantom{(x_0, y_0, z_0) =}
            \end{array} \right\{ & \begin{array}{l}
                (0, 0, 0), \\
                (\sqrt{\beta (\rho - 1)}, \sqrt{\beta (\rho - 1)}, \rho - 1), \\
                (-\sqrt{\beta (\rho - 1)}, -\sqrt{\beta (\rho - 1)}, \rho - 1)
            \end{array}
        \end{align}
        
        To study the local stability of these equilibrium points and understand the system's behavior near these points, we linearize the Lorenz system. Linearization involves approximating the system near the equilibrium points with a linear system. This is achieved by computing the Jacobian matrix $\hat{J}$ of the system at these points, given by:
        
        \begin{equation}
        \label{jacobian}
            \hat{J}= \begin{pmatrix}
            -\sigma & \sigma & 0 \\
            \rho - z_0 & -1 & -x_0 \\
            y_0 & x_0 & -\beta
            \end{pmatrix}.
        \end{equation}

        This linearized system provides crucial insights into the local dynamics around these points. Specifically, the eigenvalues of the Jacobian at these points can indicate whether the system exhibits stable, unstable, or saddle point characteristics at these locations. The behavior of the system near these equilibrium points forms the basis of our understanding of deterministic chaos in weather and climate models.

    \subsection{\label{VQE}Variational Quantum Eigensolver}
        The Variational Quantum Eigensolver is a cornerstone of quantum computing, particularly effective in the realm of quantum simulations. It is a prominent example of a hybrid quantum-classical algorithm, leveraging the strengths of both computing paradigms to address complex problems, especially those intractable in classical systems.

        VQE operates through a synergistic process involving quantum and classical computing components:
        
        \begin{itemize}
            \item \textbf{Quantum Routine:} The algorithm initiates with the preparation of a trial quantum state, $|\Psi(\theta)\rangle$, on a quantum processor. This state is parameterized by a set of variables $\theta$, which are adjustable. The quantum system then measures the expectation value of the Hamiltonian, $\langle\Psi(\theta)|\hat{H}|\Psi(\theta)\rangle$, which essentially represents the energy of the system under the given state. 

            \item \textbf{Classical Optimization:} The output from the quantum routine, i.e. the energy expectation value, is then optimized using classical algorithms. The optimization process iteratively adjusts the parameters $\theta$ to find the state that minimizes the energy, effectively seeking the system's ground state energy.
        \end{itemize}
        
    The efficiency of VQE hinges on its ability to reduce the problem of finding the ground state energy of a system to an optimization problem. This methodology has proven particularly advantageous in quantum chemistry for calculating the ground state energies of molecules, where direct computation on classical systems is exceedingly challenging due to the exponential scaling of the required resources \cite{Kandala2017, OMalley2016}.
    
    VQE's hybrid structure makes it well-suited for implementation on current and near-term quantum computers, which are often limited by factors such as qubit count and coherence times \cite{Preskill2018}. This quality positions VQE as a significant tool for advancing the field of quantum simulations, even within the constraints of existing quantum technology.

    The adaptability and versatility of VQE have broad implications in the field of quantum simulations. Beyond its established role in quantum chemistry, VQE's methodology is applicable to a range of quantum systems where the understanding of ground state properties is crucial. Its capability to handle Hamiltonians of varying complexity, from simple molecular systems to more intricate quantum models, demonstrates its potential as a versatile tool in quantum research \cite{McArdle2020, Cao2019}.

\section{\label{Methodology}Methodology}
    \subsection{Hermitianization of the Jacobian}
        The study of the Lorenz system, a set of nonlinear differential equations, is pivotal in understanding atmospheric dynamics due to its role in modeling atmospheric convection patterns. The linearized Lorenz system, represented by the $3\times3$ Jacobian of \autoref{jacobian}, inherently non-Hermitian, poses a challenge for quantum computational approaches, particularly the Variational Quantum Eigensolver, which requires a Hermitian matrix. To circumvent this, we modify the Jacobian into a Hermitian form suitable for VQE analysis.

        The approach involves constructing a Hermitian matrix H, defined as 

        \begin{equation}
            \hat{H}=(\hat{J}-\epsilon)^\dagger(\hat{J}-\epsilon),
        \end{equation}
        $\epsilon\ \in \mathbb{C}$. This transformation ensures that $\hat{H}$ is not only Hermitian but also semi-positive definite. The semi-positive definiteness of $\hat{H}$ guarantees that expectation value $\langle\Psi|\hat{H}|\Psi\rangle$ of $\hat{H}$ with a vector $|\Psi\rangle$ is zero if and only if $\epsilon$ is an eigenvalue of $\hat{J}$ and $|\Psi\rangle$ is the corresponding eigenvector, as shown in \cite{xie2023}. The analysis of equilibrium points in the Lorenz system now translates to minimizing the expectation value of $\hat{H}$.

    \subsection{Dimensionality Augmentation of the Hamiltonian}
        Quantum computing requires Hamiltonians to represented as $2^n\times2^n$ matrices, where n is the number of qubits. In our case, the Hamiltonian $\hat{H}$, being a $3\times3$ matrix, does not fit this criterion directly. To address this, we augment $\hat{H}$ into a $4\times4$ matrix by adding an additional row and column of zeros. Furthermore, to maintain the physical relevance of the augmented matrix, we insert a large diagonal term $d$ in the newly added row and column:
        \begin{equation}
            \hat{H}_{4x4} = 
            \begin{pmatrix}
                H_{11} & H_{12} & H_{13} & 0 \\
                H_{21} & H_{22} & H_{23} & 0 \\
                H_{31} & H_{32} & H_{33} & 0 \\
                0 & 0 & 0 & d
            \end{pmatrix}.
        \end{equation}
        This inclusion is crucial, as setting $d$ to zero would impede our ability to accurately determine the eigenvalues of the Jacobian matrix $\hat{J}$. If $d$ was also set to zero, the expectation value of the $4\times4$ Hamiltonian $\hat{H}$ would be zero for some vectors, irrespective of whether $\epsilon$ is an eigenvalue of $\hat{J}$ or not. This would lead to erroneous identifications of eigenvalues. Therefore, to avoid this interference and ensure the integrity of our eigenvalue search, the term $d$ must be sufficiently large. 

    \subsection{Representation of Statevector $|\Psi(\boldsymbol{\theta})\rangle$} 
        The fact that the Hamiltonian is a $4\times4$ matrix in this case means that the system can be represented in a two-qubit space ($4=2^2$). To manipulate and explore the state space, universal two-qubit gates are employed to parametrize $|\Psi(\boldsymbol{\theta})\rangle$ \cite{Vatan2004, Shende2004}. The universality of these gates means that they can create any possible two-qubit state from an initial state, thus allowing for the full exploration of the Hamiltonian's state space.

        \subsection{Gradient Descent}
            We use Gradient Descent \cite{Fletcher2000} in two distinct aspects of our study:

            \begin{enumerate}
                \item \textbf{Optimization of the Statevector:} The primary application of Gradient Descent in our methodology is to optimize the statevector $|\Psi(\boldsymbol{\theta})\rangle$. This process involves iteratively adjusting the parameters $\boldsymbol{\theta}$ of the universal two-qubit gate to minimize the expectation value of the Hamiltonian $\hat{H}$.

                \item \textbf{Identification of Eigenvalues of $\hat{J}$:} Another application of Gradient Descent in our methodology is its use in finding the specific value of $\epsilon$ which makes the expectation value of $\hat{H}$ equal to zero. When the expectation value of $\hat{H}$ reaches zero, $\epsilon$ is an eigenvalue of the Jacobian matrix $\hat{J}$.
            \end{enumerate}

        \subsection{Identification of the Eigenvalues}
            We are focusing on identifying exactly three eigenvalues corresponding to the size of the original $3\times3$ Jacobian matrix. In our larger $4\times4$ matrix, the fourth eigenvalue $d$ is just an added element and not part of the spectrum we need.
            
            Grid Search \cite{Geron2019} is employed to explore various starting points for $\epsilon$ in the Gradient Descent process. The rationale behind using Grid Search is to ensure that we capture the entire spectrum of eigenvalues of $\hat{J}$ rather than converging to a single eigenvalue. 
            
            Since the eigenvalues of real matrices appear in conjugate pairs, we can limit our search to the positive half of the imaginary axis. This approach helps reduce our search space, making the process more efficient. The process is terminated as soon as we identify the three distinct eigenvalues.

\section{Results}

    \begin{figure*}[tp]
        \includegraphics[width=1.0\linewidth]{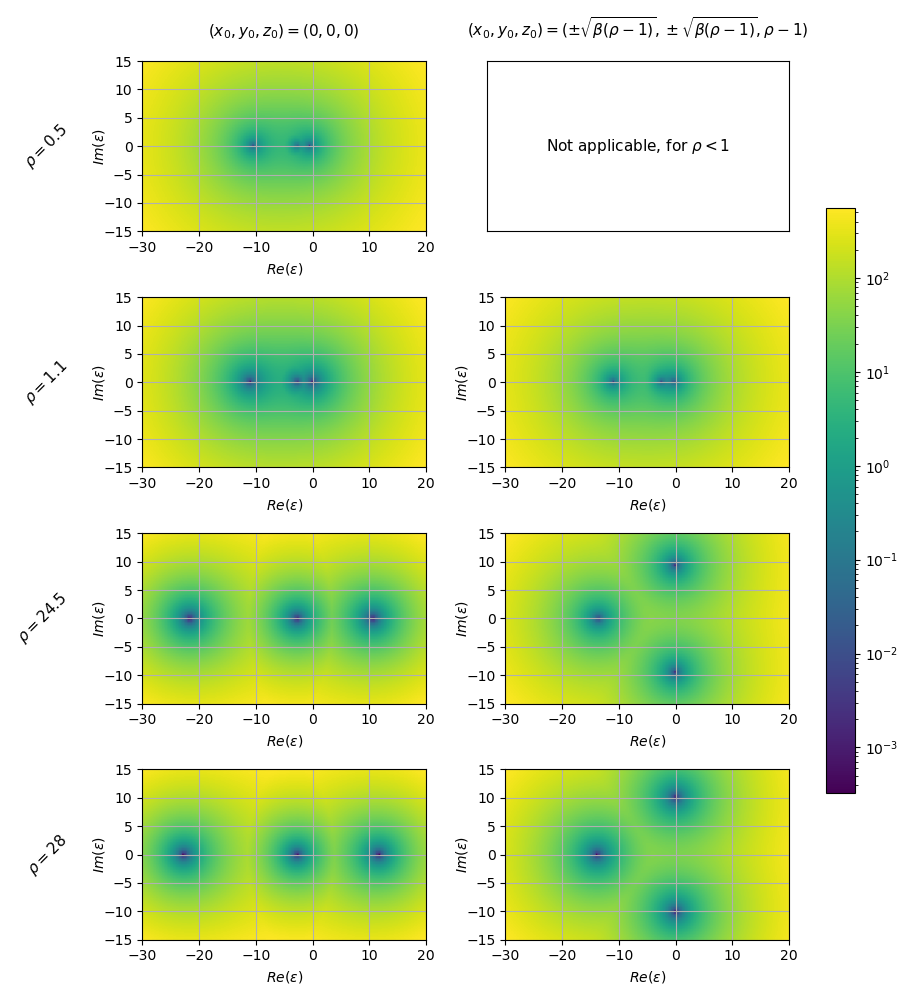}
        
        \captionsetup{width=.9\linewidth,justification=centerlast,singlelinecheck=off}
        \caption{\small\textbf{Spectrum Visualization of Equilibrium Points.} This collection of heatmaps visualizes the expectation value of the Hamiltonian $\langle \Psi_{min}|\hat{H}(\epsilon)|\Psi_{min}\rangle$ at the equilibrium points of the Lorenz system for various values of $\rho$. The expectation value is presented as a function of $\epsilon$. $|\Psi_{min}\rangle$ represents the value of $|\Psi(\theta)\rangle$ that minimizes the expectation value of $\hat{H}$ for the given $\epsilon$. The real and imaginary parts of $\epsilon$ form the axes of the plots. The expectation value becomes zero when $\epsilon$ is an eigenvalue of the Jacobian $\hat{J}$. For $\rho=0.5$, the eigenvalues corresponding to the trivial equilibrium point (0,0,0) are $-10.52, -2.67, -0.48$. For $\rho=1.1$, the eigenvalues are $-11.09, -2.67, 0.09$ (trivial) and $-11.03, -2.44, -0.20$ (non-trivial). For $\rho=24.5$, the eigenvalues are $-21.79, -2.67, 10.79$ (trivial) and $-13.65, -0.01\pm9.58i$ (non-trivial). For $\rho=28$, the eigenvalues are $-22.83, -2.67, 11.83$ (trivial) and $-13.85, 0.09\pm10.19i$ (non-trivial).}
        \label{fig:heatmap}
    \end{figure*}

    In our investigation of the Lorenz system, we focus on examining the behavior of the system under various settings of the parameter $\rho$ while maintaining constant values for the other Lorenz parameters ($\sigma = 10$, $\beta = \frac{8}{3}$). It is noteworthy that the common values for the Lorenz system are typically set at $\sigma = 10$, $\beta = \frac{8}{3}$, and $\rho = 28$, known for leading to the system's characteristic chaotic behavior. However, in this study, we explore a range of values for $\rho$ to demonstrate the distinct dynamical behaviors that emerge at each value, thereby highlighting the system's sensitivity to changes in $\rho$.

    For various values of $\rho$, we analyze the equilibrium points of the Lorenz system and compute the corresponding eigenvalues. These eigenvalues are crucial for understanding the stability and dynamical properties of the system. Below, we present our findings for each considered value of $\rho$:
    
    \begin{itemize}
        \item For $\rho = 0.5$, which is below the critical value of 1, there is only a trivial equilibrium point at $(0, 0, 0)$. The eigenvalues for this point are $-10.52$, $-2.67$, and $-0.48$, indicating a stable node.
    
        \item For $\rho = 1.1$, slightly above the critical value, the trivial equilibrium point has eigenvalues of $-11.09$, $-2.67$, and $0.09$. The positive eigenvalue indicates an unstable saddle point. Additionally, the non-trivial equilibrium points show eigenvalues of $-11.03$, $-2.44$, and $-0.20$, suggesting these points are stable nodes.
        
        \item At $\rho = 24.5$ the trivial equilibrium point has eigenvalues of $-21.79$, $-2.67$, and $10.79$, indicating an unstable saddle point. The non-trivial points have eigenvalues of $-13.65$ and $-0.01 \pm 9.58i$. The presence of complex conjugate eigenvalues with a negative real part suggests a stable spiral point.
        
        \item For $\rho = 28$, corresponding to the classical Lorenz attractor, the trivial equilibrium point has eigenvalues of $-22.83$, $-2.67$, and $11.83$, indicating an unstable saddle point. The non-trivial points have eigenvalues of $-13.85$ and $0.09 \pm 10.19i$, suggesting an unstable spiral point due to the positive real part of the complex conjugate eigenvalues.
    \end{itemize}
    
    The behavior of the Lorenz system around these equilibrium points is further illustrated in the heatmaps of \autoref{fig:heatmap}. These heatmaps display the minimum expectation value of the Hamiltonian $\langle \Psi_{min}|\hat{H}(\epsilon)|\Psi_{min}\rangle$ over a range of complex $\epsilon$ values, with the real and imaginary parts of $\epsilon$ plotted on the horizontal and vertical axes, respectively. In these visualizations, $|\Psi_{min}\rangle$ denotes the state vector that minimizes the Hamiltonian's expectation value for a given $\epsilon$. The color scale in each heatmap indicates the magnitude of the expectation value, highlighting the eigenvalues of the Jacobian matrix $\hat{J}$.

\section{Conclusions}
    The Variational Quantum Eigensolver algorithm, traditionally focused on Hermitian matrices within quantum chemistry, now extends its reach to non-Hermitian matrices. This expansion involves defining Hermitian proxies to the properties of the non-Hermitian matrices. The numerical processes benefiting from this advancement (assuming an arbitrary matrix $\hat{M}$) include:
    
    \begin{itemize}
        \item \textbf{Diagonalization}: Utilizing a Hermitian proxy,  $ \hat{H} = (\hat{M}-\epsilon)^\dagger(\hat{M}-\epsilon) $, this study demonstrates VQE's capability to compute the eigenvalues of $ \hat{M} $. This method has applications in diverse fields such as Atmospheric Physics, Structural Engineering, and Data Science.

        \item \textbf{Singular Value Decomposition (SVD)}: The application of VQE techniques for Singular Value Decomposition has been successfully demonstrated in recent studies \cite{Wang2021}. In the context of this research, we propose an alternate approach to SVD using VQE. This involves the creation of a Hermitian proxy, $ \hat{H}' = \hat{M}^\dagger \hat{M} $. By using this proxy, VQE can effectively calculate the singular values of the matrix $ \hat{M} $, since these values correspond to the eigenvalues of $ \hat{H}' $. SVD plays a crucial role in a wide range of scientific fields, including Data Analysis, Signal Processing, and Image Processing, and this method presents a novel way to leverage quantum computing for these applications.
    \end{itemize}

    In conclusion, the ability of VQE to handle non-Hermitian matrices and assist in operations like Diagonalization or SVD has an impact on various fields that rely on matrix operations. This quantum computational technique has the potential to provide faster and more efficient solutions for computationally intensive tasks, significantly extending beyond its traditional applications in quantum computing.

    However, it is important to acknowledge the current limitations of VQE, particularly in its practical implementation in atmospheric physics and other complex fields. At present, VQE's effectiveness is constrained by factors such as computational speed and error rates, especially when compared to established conventional methods. Despite these challenges, the rapidly evolving landscape of quantum computing holds promise for the advancement of VQE's efficiency and robustness.

    The application of VQE to non-Hermitian matrices in atmospheric physics is especially promising. As quantum computing technology continues to advance, it is anticipated that VQE and similar quantum algorithms will play a crucial role in enhancing simulations and analyses of atmospheric phenomena. This could include significant improvements in the simulations and analyses of atmospheric phenomena, particularly in the realms of weather and climate dynamics. The ongoing research in this domain points towards a transformative impact of quantum computing in atmospheric physics, providing new methodologies to address the complexities inherent in atmospheric systems.
    
    In summary, while VQE currently faces limitations in practical applications, its potential to revolutionize computational methods, especially in fields reliant on complex matrix operations, is substantial. The future of quantum computing in atmospheric physics and other domains is poised to bring about significant advancements, with VQE at the forefront of this technological evolution.

\bibliography{bib}

\begin{thebibliography}{16}
\expandafter\ifx\csname natexlab\endcsname\relax\def\natexlab#1{#1}\fi
\expandafter\ifx\csname bibnamefont\endcsname\relax
  \def\bibnamefont#1{#1}\fi
\expandafter\ifx\csname bibfnamefont\endcsname\relax
  \def\bibfnamefont#1{#1}\fi
\expandafter\ifx\csname citenamefont\endcsname\relax
  \def\citenamefont#1{#1}\fi
\expandafter\ifx\csname url\endcsname\relax
  \def\url#1{\texttt{#1}}\fi
\expandafter\ifx\csname urlprefix\endcsname\relax\def\urlprefix{URL }\fi
\providecommand{\bibinfo}[2]{#2}
\providecommand{\eprint}[2][]{\url{#2}}

\bibitem[{\citenamefont{McArdle et~al.}(2020)\citenamefont{McArdle, Endo, Aspuru-Guzik, Benjamin, and Yuan}}]{McArdle2020}
\bibinfo{author}{\bibfnamefont{S.}~\bibnamefont{McArdle}}, \bibinfo{author}{\bibfnamefont{S.}~\bibnamefont{Endo}}, \bibinfo{author}{\bibfnamefont{A.}~\bibnamefont{Aspuru-Guzik}}, \bibinfo{author}{\bibfnamefont{S.~C.} \bibnamefont{Benjamin}}, \bibnamefont{and} \bibinfo{author}{\bibfnamefont{X.}~\bibnamefont{Yuan}}, \bibinfo{journal}{Rev. Mod. Phys.} \textbf{\bibinfo{volume}{92}}, \bibinfo{pages}{015003} (\bibinfo{year}{2020}), \urlprefix\url{https://link.aps.org/doi/10.1103/RevModPhys.92.015003}.

\bibitem[{\citenamefont{Armaos et~al.}(2020)\citenamefont{Armaos, Badounas, Deligiannis, and Lianos}}]{Armaos2020}
\bibinfo{author}{\bibfnamefont{V.}~\bibnamefont{Armaos}}, \bibinfo{author}{\bibfnamefont{D.~A.} \bibnamefont{Badounas}}, \bibinfo{author}{\bibfnamefont{P.}~\bibnamefont{Deligiannis}}, \bibnamefont{and} \bibinfo{author}{\bibfnamefont{K.}~\bibnamefont{Lianos}}, \bibinfo{journal}{Applied Physics A} \textbf{\bibinfo{volume}{126}}, \bibinfo{pages}{625} (\bibinfo{year}{2020}), ISSN \bibinfo{issn}{1432-0630}, \urlprefix\url{https://doi.org/10.1007/s00339-020-03755-4}.

\bibitem[{\citenamefont{Tennie and Palmer}(2023)}]{Tennie2023}
\bibinfo{author}{\bibfnamefont{F.}~\bibnamefont{Tennie}} \bibnamefont{and} \bibinfo{author}{\bibfnamefont{T.~N.} \bibnamefont{Palmer}}, \bibinfo{journal}{Bulletin of the American Meteorological Society} \textbf{\bibinfo{volume}{104}}, \bibinfo{pages}{E488 } (\bibinfo{year}{2023}), \urlprefix\url{https://journals.ametsoc.org/view/journals/bams/104/2/BAMS-D-22-0031.1.xml}.

\bibitem[{\citenamefont{Peruzzo et~al.}(2014)\citenamefont{Peruzzo, McClean, Shadbolt, Yung, Zhou, Love, Aspuru-Guzik, and O'Brien}}]{Peruzzo2014}
\bibinfo{author}{\bibfnamefont{A.}~\bibnamefont{Peruzzo}}, \bibinfo{author}{\bibfnamefont{J.}~\bibnamefont{McClean}}, \bibinfo{author}{\bibfnamefont{P.}~\bibnamefont{Shadbolt}}, \bibinfo{author}{\bibfnamefont{M.-H.} \bibnamefont{Yung}}, \bibinfo{author}{\bibfnamefont{X.-Q.} \bibnamefont{Zhou}}, \bibinfo{author}{\bibfnamefont{P.~J.} \bibnamefont{Love}}, \bibinfo{author}{\bibfnamefont{A.}~\bibnamefont{Aspuru-Guzik}}, \bibnamefont{and} \bibinfo{author}{\bibfnamefont{J.~L.} \bibnamefont{O'Brien}}, \bibinfo{journal}{Nature Communications} \textbf{\bibinfo{volume}{5}}, \bibinfo{pages}{4213} (\bibinfo{year}{2014}), ISSN \bibinfo{issn}{2041-1723}, \urlprefix\url{https://doi.org/10.1038/ncomms5213}.

\bibitem[{\citenamefont{McClean et~al.}(2016)\citenamefont{McClean, Romero, Babbush, and Aspuru-Guzik}}]{McClean2016}
\bibinfo{author}{\bibfnamefont{J.~R.} \bibnamefont{McClean}}, \bibinfo{author}{\bibfnamefont{J.}~\bibnamefont{Romero}}, \bibinfo{author}{\bibfnamefont{R.}~\bibnamefont{Babbush}}, \bibnamefont{and} \bibinfo{author}{\bibfnamefont{A.}~\bibnamefont{Aspuru-Guzik}}, \bibinfo{journal}{New Journal of Physics} \textbf{\bibinfo{volume}{18}}, \bibinfo{pages}{023023} (\bibinfo{year}{2016}), \urlprefix\url{https://dx.doi.org/10.1088/1367-2630/18/2/023023}.

\bibitem[{\citenamefont{Lorenz}(1963)}]{Lorentz1963}
\bibinfo{author}{\bibfnamefont{E.~N.} \bibnamefont{Lorenz}}, \bibinfo{journal}{Journal of Atmospheric Sciences} \textbf{\bibinfo{volume}{20}}, \bibinfo{pages}{130 } (\bibinfo{year}{1963}), \urlprefix\url{https://journals.ametsoc.org/view/journals/atsc/20/2/1520-0469_1963_020_0130_dnf_2_0_co_2.xml}.

\bibitem[{\citenamefont{Kandala et~al.}(2017)\citenamefont{Kandala, Mezzacapo, Temme, Takita, Brink, Chow, and Gambetta}}]{Kandala2017}
\bibinfo{author}{\bibfnamefont{A.}~\bibnamefont{Kandala}}, \bibinfo{author}{\bibfnamefont{A.}~\bibnamefont{Mezzacapo}}, \bibinfo{author}{\bibfnamefont{K.}~\bibnamefont{Temme}}, \bibinfo{author}{\bibfnamefont{M.}~\bibnamefont{Takita}}, \bibinfo{author}{\bibfnamefont{M.}~\bibnamefont{Brink}}, \bibinfo{author}{\bibfnamefont{J.~M.} \bibnamefont{Chow}}, \bibnamefont{and} \bibinfo{author}{\bibfnamefont{J.~M.} \bibnamefont{Gambetta}}, \bibinfo{journal}{Nature} \textbf{\bibinfo{volume}{549}}, \bibinfo{pages}{242} (\bibinfo{year}{2017}).

\bibitem[{\citenamefont{O'Malley et~al.}(2016)\citenamefont{O'Malley, Babbush, Kivlichan, Romero, McClean, Barends, Kelly, Roushan, Tranter, Ding et~al.}}]{OMalley2016}
\bibinfo{author}{\bibfnamefont{P.~J.~J.} \bibnamefont{O'Malley}}, \bibinfo{author}{\bibfnamefont{R.}~\bibnamefont{Babbush}}, \bibinfo{author}{\bibfnamefont{I.~D.} \bibnamefont{Kivlichan}}, \bibinfo{author}{\bibfnamefont{J.}~\bibnamefont{Romero}}, \bibinfo{author}{\bibfnamefont{J.~R.} \bibnamefont{McClean}}, \bibinfo{author}{\bibfnamefont{R.}~\bibnamefont{Barends}}, \bibinfo{author}{\bibfnamefont{J.}~\bibnamefont{Kelly}}, \bibinfo{author}{\bibfnamefont{P.}~\bibnamefont{Roushan}}, \bibinfo{author}{\bibfnamefont{A.}~\bibnamefont{Tranter}}, \bibinfo{author}{\bibfnamefont{N.}~\bibnamefont{Ding}}, \bibnamefont{et~al.}, \bibinfo{journal}{Phys. Rev. X} \textbf{\bibinfo{volume}{6}}, \bibinfo{pages}{031007} (\bibinfo{year}{2016}), \urlprefix\url{https://link.aps.org/doi/10.1103/PhysRevX.6.031007}.

\bibitem[{\citenamefont{Preskill}(2018)}]{Preskill2018}
\bibinfo{author}{\bibfnamefont{J.}~\bibnamefont{Preskill}}, \bibinfo{journal}{{Quantum}} \textbf{\bibinfo{volume}{2}}, \bibinfo{pages}{79} (\bibinfo{year}{2018}), ISSN \bibinfo{issn}{2521-327X}, \urlprefix\url{https://doi.org/10.22331/q-2018-08-06-79}.

\bibitem[{\citenamefont{Cao et~al.}(2019)\citenamefont{Cao, Romero, Olson, Degroote, Johnson, Kieferová, Kivlichan, Menke, Peropadre, Sawaya et~al.}}]{Cao2019}
\bibinfo{author}{\bibfnamefont{Y.}~\bibnamefont{Cao}}, \bibinfo{author}{\bibfnamefont{J.}~\bibnamefont{Romero}}, \bibinfo{author}{\bibfnamefont{J.~P.} \bibnamefont{Olson}}, \bibinfo{author}{\bibfnamefont{M.}~\bibnamefont{Degroote}}, \bibinfo{author}{\bibfnamefont{P.~D.} \bibnamefont{Johnson}}, \bibinfo{author}{\bibfnamefont{M.}~\bibnamefont{Kieferová}}, \bibinfo{author}{\bibfnamefont{I.~D.} \bibnamefont{Kivlichan}}, \bibinfo{author}{\bibfnamefont{T.}~\bibnamefont{Menke}}, \bibinfo{author}{\bibfnamefont{B.}~\bibnamefont{Peropadre}}, \bibinfo{author}{\bibfnamefont{N.~P.~D.} \bibnamefont{Sawaya}}, \bibnamefont{et~al.}, \bibinfo{journal}{Chemical Reviews} \textbf{\bibinfo{volume}{119}}, \bibinfo{pages}{10856} (\bibinfo{year}{2019}), \bibinfo{note}{pMID: 31469277}, \eprint{https://doi.org/10.1021/acs.chemrev.8b00803}, \urlprefix\url{https://doi.org/10.1021/acs.chemrev.8b00803}.

\bibitem[{\citenamefont{Xie et~al.}(2023)\citenamefont{Xie, Xue, and Zhang}}]{xie2023}
\bibinfo{author}{\bibfnamefont{X.-D.} \bibnamefont{Xie}}, \bibinfo{author}{\bibfnamefont{Z.-Y.} \bibnamefont{Xue}}, \bibnamefont{and} \bibinfo{author}{\bibfnamefont{D.-B.} \bibnamefont{Zhang}}, \emph{\bibinfo{title}{Variational quantum eigensolvers for the non-hermitian systems by variance minimization}} (\bibinfo{year}{2023}), \eprint{2305.19807}.

\bibitem[{\citenamefont{Vatan and Williams}(2004)}]{Vatan2004}
\bibinfo{author}{\bibfnamefont{F.}~\bibnamefont{Vatan}} \bibnamefont{and} \bibinfo{author}{\bibfnamefont{C.}~\bibnamefont{Williams}}, \bibinfo{journal}{Phys. Rev. A} \textbf{\bibinfo{volume}{69}}, \bibinfo{pages}{032315} (\bibinfo{year}{2004}), \urlprefix\url{https://link.aps.org/doi/10.1103/PhysRevA.69.032315}.

\bibitem[{\citenamefont{Shende et~al.}(2004)\citenamefont{Shende, Markov, and Bullock}}]{Shende2004}
\bibinfo{author}{\bibfnamefont{V.~V.} \bibnamefont{Shende}}, \bibinfo{author}{\bibfnamefont{I.~L.} \bibnamefont{Markov}}, \bibnamefont{and} \bibinfo{author}{\bibfnamefont{S.~S.} \bibnamefont{Bullock}}, \bibinfo{journal}{Phys. Rev. A} \textbf{\bibinfo{volume}{69}}, \bibinfo{pages}{062321} (\bibinfo{year}{2004}), \urlprefix\url{https://link.aps.org/doi/10.1103/PhysRevA.69.062321}.

\bibitem[{\citenamefont{Fletcher}(2000)}]{Fletcher2000}
\bibinfo{author}{\bibfnamefont{R.}~\bibnamefont{Fletcher}} (\bibinfo{publisher}{John Wiley \& Sons, Ltd}, \bibinfo{year}{2000}), ISBN \bibinfo{isbn}{9781118723203}, \urlprefix\url{https://onlinelibrary.wiley.com/doi/abs/10.1002/9781118723203.ch1}.

\bibitem[{\citenamefont{Geron}(2019)}]{Geron2019}
\bibinfo{author}{\bibfnamefont{A.}~\bibnamefont{Geron}}, \emph{\bibinfo{title}{Hands-on machine learning with scikit-learn, keras, and {TensorFlow}}} (\bibinfo{publisher}{O'Reilly Media}, \bibinfo{address}{Sebastopol, CA}, \bibinfo{year}{2019}), \bibinfo{edition}{2nd} ed.

\bibitem[{\citenamefont{Wang et~al.}(2021)\citenamefont{Wang, Song, and Wang}}]{Wang2021}
\bibinfo{author}{\bibfnamefont{X.}~\bibnamefont{Wang}}, \bibinfo{author}{\bibfnamefont{Z.}~\bibnamefont{Song}}, \bibnamefont{and} \bibinfo{author}{\bibfnamefont{Y.}~\bibnamefont{Wang}}, \bibinfo{journal}{{Quantum}} \textbf{\bibinfo{volume}{5}}, \bibinfo{pages}{483} (\bibinfo{year}{2021}), ISSN \bibinfo{issn}{2521-327X}, \urlprefix\url{https://doi.org/10.22331/q-2021-06-29-483}.

\end{thebibliography}
\end{document}